\documentclass[aps,twocolumn,amsmath,amssymb,floatfix,superscriptaddress,nofootinbib]{revtex4-2}

\usepackage{graphicx}
\usepackage{dcolumn}
\usepackage{bm}
\usepackage{latexsym}
\usepackage{color}
\usepackage{xcolor}
\usepackage[colorlinks=true,linkcolor=blue,citecolor=blue]{hyperref}

\begin{document}

\title{Particle acceleration at magnetized, relativistic turbulent shock fronts}

\author{Virginia Bresci} 
\affiliation{Institut d'Astrophysique de Paris,
CNRS -- Sorbonne Universit\'e,
98 bis boulevard Arago, F-75014 Paris, France}
\affiliation{CEA, DAM, DIF, F-91297 Arpajon, France}
\affiliation{Leibniz-Institut für Astrophysik Potsdam (AIP), An der Sternwarte 16, 14482 Potsdam, Germany}
\author{Martin Lemoine} 
\affiliation{Institut d'Astrophysique de Paris,
CNRS -- Sorbonne Universit\'e,
98 bis boulevard Arago, F-75014 Paris, France}
\author{Laurent Gremillet} 
\affiliation{CEA, DAM, DIF, F-91297 Arpajon, France}
\affiliation{Universit\'e Paris-Saclay, CEA, LMCE, 91680 Bruy\`eres-le-Ch\^atel, France}

\date{\today}

\begin{abstract} 
The efficiency of particle acceleration at shock waves in relativistic, magnetized astrophysical outflows is a debated topic with far-reaching implications. Here, we study the impact of well-developed turbulence in the pre-shock plasma. Our simulations demonstrate that, for a mildly relativistic, magnetized pair shock (Lorentz factor $\gamma_{\rm sh} \simeq 2.7$, magnetization level $\sigma \simeq 0.01$), strong turbulence can revive particle acceleration in a superluminal configuration that otherwise prohibits it. Depending on the initial plasma temperature and magnetization, shock-drift or diffusive-type acceleration governs particle energization, producing powerlaw spectra $\mathrm{d}N/\mathrm{d}\gamma \propto \gamma^{-s}$ with $s \sim 2.5-3.5$. At larger magnetization levels, stochastic acceleration within the pre-shock turbulence becomes competitive and can even take over shock acceleration.
\end{abstract}

\maketitle

\section{Introduction} 

The nonthermal radiative spectra observed from high-energy, relativistic astrophysical sources point to a bulk energy reservoir being dissipated into accelerated particles through, \textit{e.g.}, magnetic reconnection \citep{Guo_2014,2015SSRv..191..545K,Werner_2017, 2019ApJ...880...37P}, shock acceleration \cite{2008ApJ...682L...5S,Sironi_2015}, or turbulent Fermi processes \cite{Zhdankin+_17, CS_18,2019PhRvL.122e5101Z,Bresci+22}. As direct offsprings of the powerful outflows associated with those sources, collisionless shock waves emerge as natural dissipation agents \cite{2012SSRv..173..309B}.
Yet, in relativistic and magnetized plasmas, particle acceleration appears inhibited by the generic superluminal nature of the shock  \citep{Begelman_1990, Lemoine_2006, Niemiec+_06, 2009MNRAS.393..587P, 2009ApJ...698.1523S, Lemoine_2010, Sironi_2013}. Specifically, \emph{ab initio} particle-in-cell (PIC) numerical simulations have revealed a powerlaw tail of nonthermal particles increasingly shrinking as the magnetization parameter $\sigma$ \footnote{We define the magnetization level as $\sigma \equiv \langle B^2 \rangle/4\pi \langle \epsilon \rangle$, in terms of mean-squared magnetic field \unexpanded{$\langle B^2 \rangle$} and the energy density \unexpanded{$\langle \epsilon \rangle = n\langle \gamma \rangle m_{\rm e} c^2$} as measured in the simulation (downstream) frame; $n$ represents the total apparent density. When measured in the comoving turbulence frame, the resulting value of $\sigma$ may be a factor $\sim2$ larger.} rises above $\sim 10^{-4}$, until it vanishes at $\sigma \sim 10^{-2}$ \citep{Sironi_2013, Plotnikov+_18}. Given that a significant magnetization is expected in a wide class of high-energy astrophysical jets, {\it e.g.}  gamma-ray bursts, pulsar wind nebulae or blazars, this result challenges the role of shocks as sources of high-energy particles~\citep{Sironi+_15}.

One limitation of previous numerical studies based on fully PIC simulations, though, is to consider laminar inflow conditions, \emph{i.e.}, nonturbulent, homogeneous background plasmas of uniform magnetization. Notable exceptions are, to our knowledge, Ref.~\cite{2023ApJ...947L..10D}, which shows that the interaction of a shock with a monochromatic linear eigenmode of the upstream plasma leads to particle acceleration in the resultant downstream turbulence,  Refs.~\cite{2019JPhCS1400b2005R,2020JPhCS1697a2027R} which examine the impact of a superposition of magnetostatic plane waves seeded upstream, and finally Ref.~\cite{CeruttiGiacinti_2020} that considers the influence of an anisotropic transverse upstream magnetic profile. 
The presence of a strong turbulence upstream of a fast shock may change the picture in various ways:
it may preaccelerate the plasma particles via a stochastic Fermi process \citep{Zhdankin+_17,CS_18,2019PhRvL.122e5101Z,Bresci+22,2022PhRvL.129u5101L}, just as it may corrugate the shock front so that the turbulence does not transform trivially through the shock \citep{Zank+_02, Mizuno+_11, Lemoine_16, Lemoine+_16, Demidem+_18, Trotta+_21,2022ApJ...926..109N}, possibly unlocking particles from the field lines and enabling their acceleration. In light of these considerations, the paradigm of inefficient relativistic magnetized shocks as particle accelerators needs to be revisited in the likely common case of turbulent environments.


To this goal, we here report on the first PIC simulations of relativistic shocks propagating in turbulent, magnetized pair plasmas. We demonstrate that, despite a substantial magnetization ($\sigma \sim 0.01$), shock acceleration is manifest, and that the particle spectrum develops a powerlaw tail extending in time, which is absent without turbulence. This drastic change involves a significant ($\delta B>B_0$), but not too strong upstream turbulence, otherwise its own contribution to particle acceleration can supersede that of the shock. The discussion is organized as follows: we detail the simulation technique in Sec.~\ref{sec:method}, then investigate the acceleration processes at play in Sec.~\ref{sec:results}, before concluding and summarizing our results in Sec.~\ref{sec:conclusions}. 

\section{Numerical method}\label{sec:method}

We perform such simulations using the fully electromagnetic and relativistic \textsc{calder} code \citep{Calder, Lemoine_2019, Vanthieghem_22}.  Turbulence is excited close to the right-hand side of the domain, in a pair plasma continually injected along $-\boldsymbol{\hat{x}}$ at a relativistic velocity $v_\infty=-0.87\,c$ (Lorentz factor $\gamma_\infty=2$). The flow is left to propagate across the domain until the turbulence hits its left-hand side.  Switching the local boundary condition from open to reflective at that time triggers a rightward propagating shock wave~\cite{2008ApJ...682L...5S}, which sweeps the incoming turbulent plasma. The simulation frame thus corresponds to the downstream rest frame of the shock. Due to physical constraints discussed thereafter, we restrict ourselves to 2D3V geometry (2D in space, 3D in momentum). A uniform magnetic guide field $\boldsymbol{B_0}$ is applied along the (out-of-plane) $z$ direction with corresponding magnetization level $\sigma_0$.

\begin{figure*}
\includegraphics[width=0.98\textwidth]{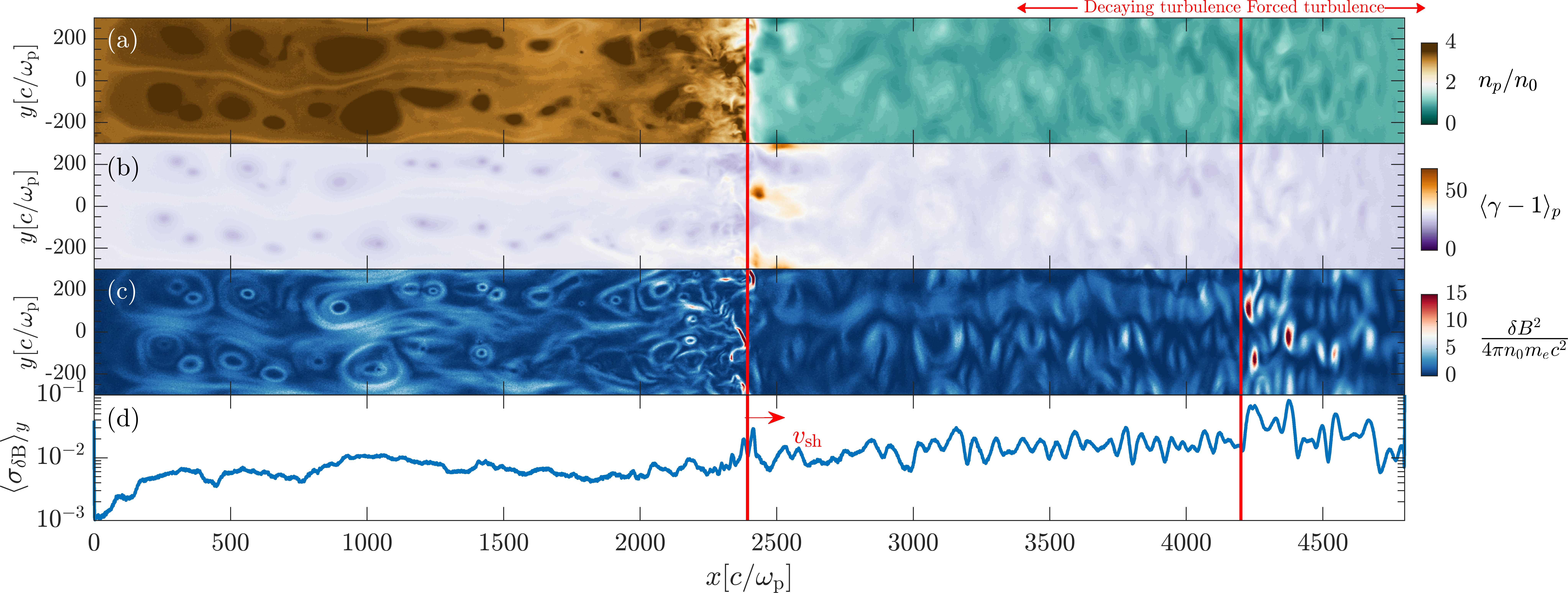}
        \caption{Spatial distributions of (a) apparent positron density $n_p$ (normalized to the apparent far-upstream positron density $n_0$), (b) mean kinetic energy per positron $\langle \gamma-1 \rangle$, and (c) squared turbulent magnetic field $\delta B_x^2 + \delta B_y^2$ (normalized to $4 \pi n_0 m_{\rm e} c^2$) for simulation S2 at $t \simeq 12\,400\, \omega_p^{-1}$. In panel (d), the longitudinal profile of the turbulent magnetization $\sigma_{\delta B} = (\delta B_x^2+\delta B_y^2)/(4\pi \sum_\alpha n_\alpha \langle \gamma \rangle_\alpha m_{\rm e}c^2)$, with $n_\alpha$ the total (apparent) density of the plasma, averaged over the transverse dimension ($y$). The left and right vertical lines locate the shock front and the boundary of forced turbulence.
        }
        \label{fig:turb_shock}
\end{figure*}

The turbulence is driven in the rest frame of the drifting plasma, in a finite region covering a few stirring length scales $\ell_{\rm c}$. Elsewhere, the energy injection in the system is halted and the turbulence is let to develop freely, thereby initiating the cascade before it interacts with the shock. Following~\cite{Zhdankin+_17}, we aim at exciting turbulence on a scale $\ell_{\rm c}$ as large as possible compared to the kinetic scale $c/\omega_{\rm p}$, in order to simulate an inertial range under near-magnetohydrodynamics conditions where the turbulence cascades down to the dissipative range. Those simulations are demanding, because the interaction between the turbulence and the shock must be followed over a long enough timespan, the transverse dimension must accommodate $\gtrsim 1-2 \,\ell_{\rm c}$,  and also because the need to stir turbulence in the plasma rest frame brings in further constraints due to time dilation effects.

In detail, the simulation domain contains $48\,000 \times 6\,000$ cells along the $x$ and $y$ axes and the simulation is run over up to $120\,000$ time steps. The mesh size is $\Delta x = \Delta y = 0.1 \,c/\omega_{\rm p}$~\footnote{$\omega_{\rm p} \equiv (4 \pi n_\infty e^2/m_{\rm e})^{1/2}$ denotes the nonrelativistic plasma frequency of the far upstream plasma, with $n_\infty = 2n_0/\gamma_\infty$ the (total) proper density and $n_0$ the apparent density of one species.} and the time step is $\Delta t = 0.99\,\Delta x/c$. Periodic boundary conditions are used for particles and fields in the transverse direction. Turbulence is excited in the interval $4200 \leq \omega_{\rm p}x/c \leq 4800$ through external magnetic perturbations $(\delta B_x,\delta B_y)$ implemented as plane waves. Those are seeded following a Langevin antenna scheme~\citep{TenBarge_2014,Zhdankin+_17, Bresci+22}, which acts in the proper plasma frame (see Appendix~\ref{sec:sim_param_turb_scheme}). The mean wavenumber $\langle k' \rangle \simeq 2.9 \times 2 \pi /L_y$ ($L_y = 600\,c/\omega_{\rm p}$ the transverse box size) implies a comoving coherence length  $\ell_{\rm c} = 2\pi/\langle k' \rangle \simeq 200\,c/\omega_{\rm p}$ (primed quantities are evaluated in the local plasma frame).  At $x\leq 4200\,c/\omega_{\rm p}$, stirring is halted, hence fluctuations evolve on a (proper) nonlinear timescale $\tau_{\rm nl}'\equiv\ell_{\rm c}/v_{\rm A} \sim 1\,200\, \omega_{\rm p}^{-1}$ ($v_{\rm A}\simeq c\sqrt{\sigma_{\delta B}} \simeq 0.17\,c$ is the Alfv\'en velocity), corresponding to $\tau_{\rm nl} =  \gamma_{\infty} \ell_{\rm c}/v_{\rm A} \sim 2\,400 \, \omega_{\rm p}^{-1}$ in the simulation frame.

We report here on three main simulations exploring different initial background magnetization levels $\sigma_0 \sim 10^{-4} \rightarrow 10^{-3}$, turbulent magnetization $\sigma_{\delta B} \sim 10^{-2}\rightarrow 10^{-1}$ and initial proper plasma temperatures $T$, from subrelativistic to relativistic (see Appendix~\ref{sec:sim_param_turb_scheme}). In detail, $\left\{\sigma_0,\,\,\sigma_{\delta B},\,\,k_{\rm B}T/m_ec^2 \right\} = \left\{ 0.2\times10^{-3},\,\, 2 \times 10^{-2},\,\,0.1 \right\}$ (hereafter S1), $\left \{ 0.6 \times 10^{-3},\,\,1\times 10^{-2},\,\,4.\right\}$ (S2), $\left \{ 0.6 \times 10^{-3},\,\, 10^{-1},\,\,4.\right\}$ (S3). 
A relativistically hot initial plasma as in S2 and S3 could describe internal shocks inside a strongly turbulent jet. We have run ancillary simulations, in particular S2a, similar to S2 albeit deprived of turbulence, S2b which retains open boundary conditions and thus models drifting turbulence without a shock, and finally S4, for which $k_{\rm B}T/m_ec^2=0.1$ as in S1, but with larger $\sigma_0$ and $\sigma_{\delta B}$, as in S2.

\section{Results} \label{sec:results}
Figure~\ref{fig:turb_shock} displays (from top to bottom) the spatial distributions of plasma positron density, mean Lorentz factor and magnetization level at the final simulation time $t \simeq 12\,400\,\omega_{\rm p}^{-1}$ for simulation S2. The rightward-moving shock front has then reached $x\simeq 2\,400\, c/\omega_{\rm p}$. Once swept up by the shock, the plasma is compressed by a factor of $\simeq 3.5$ and the mean kinetic energy per particle slightly increases, in good agreement with the shock-crossing conditions~\cite{Kirk-Duffy} [Figs.~\ref{fig:turb_shock}(a,b)]. Magnetic fluctuations are at their highest near the right boundary where turbulence is continually excited [Fig.~\ref{fig:turb_shock}(c)]. The $\sim 2\,000\, c/\omega_{\rm p}$ distance between the shock and the boundary of forced turbulence is then just below the minimum distance $c \tau_{\rm nl}$ needed for nonlinear evolution of the turbulence. This guarantees that the shock-turbulence interaction is not affected by the stirring procedure in the right part of the domain.

As shown in Fig.~\ref{fig:turb_shock}(d), the turbulence profile reaches an approximately steady state by the time it encounters the shock. We have checked that the spatial power spectrum of magnetic fluctuations in the transverse $y$ direction, which extends over three orders of magnitude, shows a general scaling $\propto k_y^{-5/3}$ at large scales $k_y \lesssim 15 \ell_c^{-1}$, and a steeper behavior at kinetic scales, consistent with previous PIC studies of nondrifting decaying turbulence \citep{CS_18}.

Further upstream, corresponding to an earlier stage in the turbulence evolution, stronger fluctuations are observed, as expected. The eddies are compressed when transiting across the shock and continue interacting until the turbulence eventually relaxes further downstream. This general picture resembles that observed in MHD simulations of the interaction of a monochromatic, linear plasma eigenmode with a relativistic shock front~\citep{Demidem+_18}. 

\begin{figure}[t]
\includegraphics[width=0.47\textwidth]{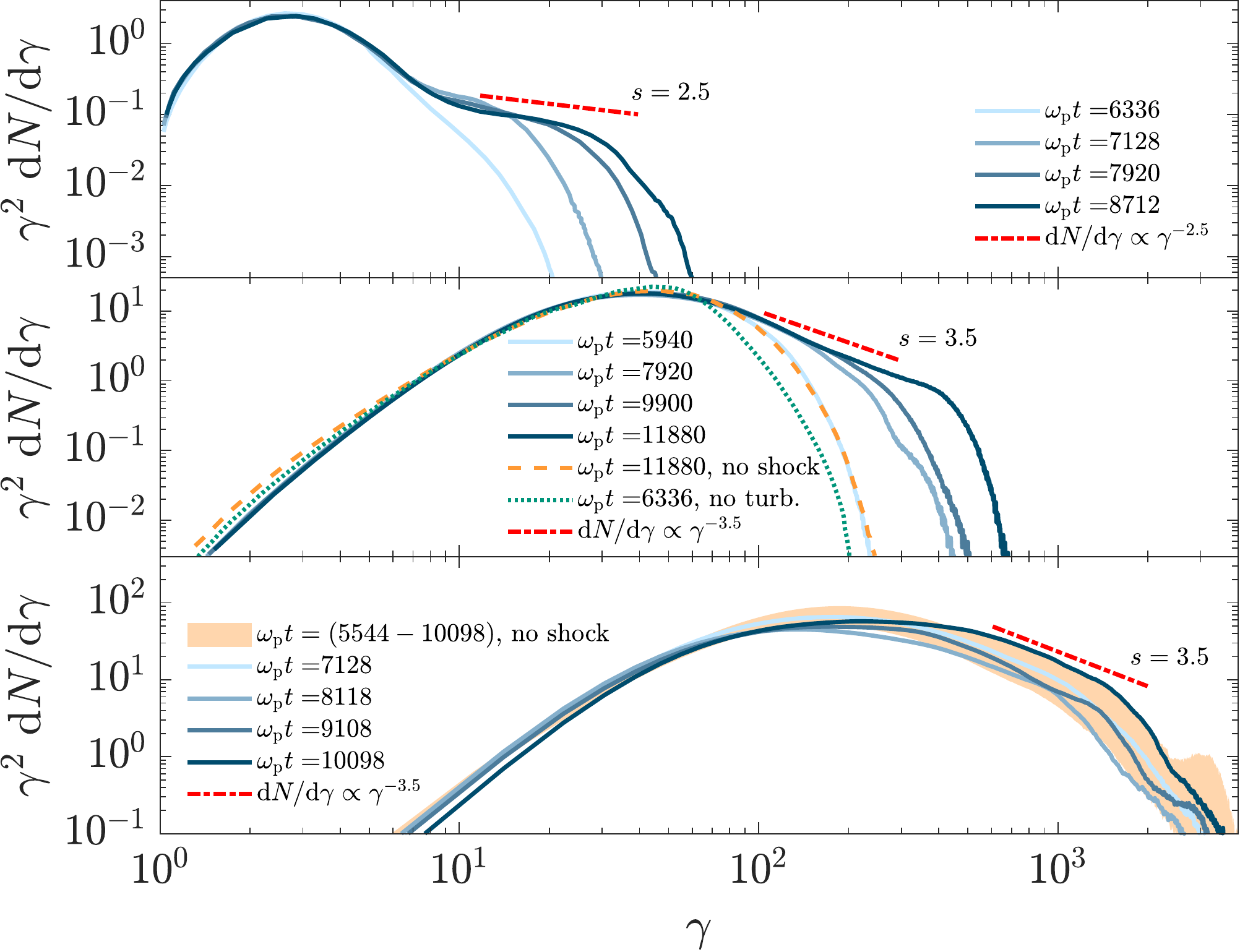}
\caption{Particle energy spectra $\gamma^2 {\rm d}N/{\rm d} \gamma$ at different times (from light to dark solid blue) in simulations S1, S2 and S3, from top to bottom. Middle panel: in dotted line, spectrum from S2a, {\it i.e.}, a shock interacting with a non-turbulent plasma, and in dashed line, spectrum from S2b, {\it i.e.}, a drifting turbulence without a shock, both in conditions otherwise similar to S2. In the lower panel, the light orange band delineates the range of spectra measured in a simulation similar to S3 albeit without a shock, as extracted at various places and times.}
\label{fig:spectrum_evolution}
\end{figure} 

The shock moves at velocity $\simeq 0.4\,c$ in the simulation (downstream) frame ($0.36\,c$ is predicted by the shock-crossing conditions \citep{Kirk-Duffy}), and consequently at $v_{\rm sh} \simeq 0.93\,c$ in the upstream frame, corresponding to a shock Lorentz factor $\gamma_{\rm sh} \simeq 2.7$. Ahead of the shock, the transversely averaged magnetic field strength is $\delta B/(m_{\rm e} \omega_{\rm p} c/e) \simeq 0.8$, so that particles with Lorentz factor $\gamma \simeq 100-300$ have a gyroradius $r_{\rm g} \simeq 120-360\,c/\omega_{\rm p}$.

Figure~\ref{fig:spectrum_evolution} plots the time evolution of the particle energy spectra $\gamma^2 {\rm d}N/{\rm d} \gamma$ (per log-interval of energy) in each of our simulations, as integrated over a moving window centered on the shock front position $x_{\rm sh}$ (located from the plasma density map) and with a $200 \,c/\omega_{\rm p}$ half-width along $x$. The peaks and widths of the spectra -- for those simulations with shock -- are consistent with shock dissipation, as predicted by the shock-crossing conditions. Remarkably, a suprathermal tail develops in all cases, with approximate powerlaw index $s\simeq 2.5\rightarrow 3.5$ (as defined through ${\rm d}N/{\rm d}\gamma\propto\gamma^{-s}$), providing manifest evidence of particle acceleration. In S1 and S2, and unlike in S3, the maximal energy is seen to increase with time.
 
\begin{figure}[t]
        \includegraphics[width=0.45\textwidth]{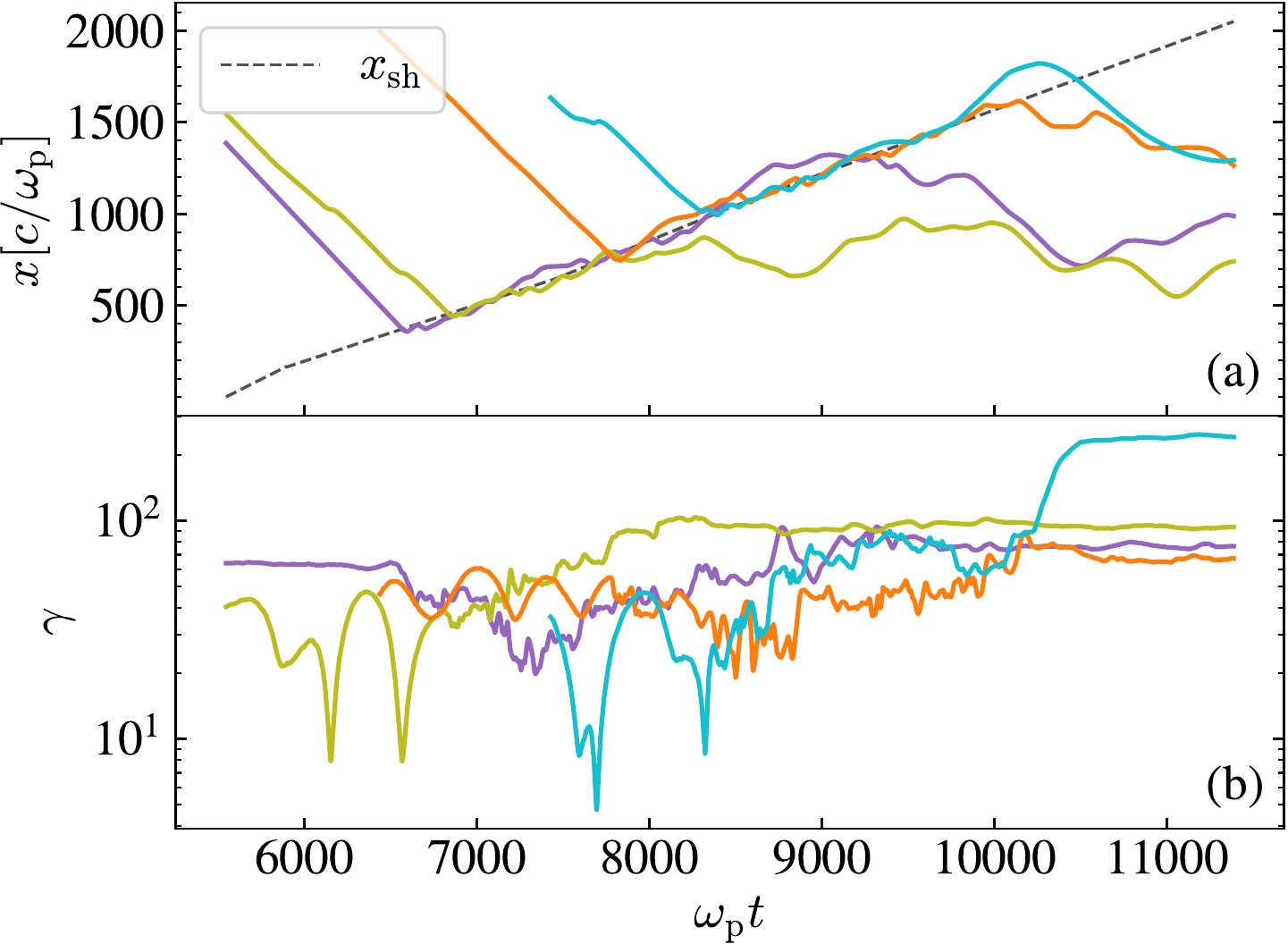}
        \caption{Position $x$ along the shock normal (a) and Lorentz factor $\gamma$ (b) versus time for $5$ particles with initial Lorentz factor $30<\gamma<100$ in simulation S2. The dashed curve in (a) indicates the shock front position $x_{\rm sh}(t)$.}  
       \label{fig: ptest}
\end{figure} 
 
The middle panel of Fig.~\ref{fig:spectrum_evolution} presents additional spectra from the turbulence-free simulation S2a, thus where the shock forms immediately in the external field $\boldsymbol{B_0}$ (dotted line), and from shock-free simulation S2b, which contains only drifting turbulence (dashed line). Clearly, the suprathermal tail only arises when the shock interacts with the turbulence. The absence of particle acceleration in S2a can be attributed to the background magnetization level $\sigma \equiv \sigma_0 \simeq 10^{-3}$, large enough to inhibit Fermi cycles around the shock. In S2b, the magnetic fluctuations are too slow to accelerate particles, as the characteristic acceleration timescale $t_{\rm acc} \sim \gamma_{\infty} \,c \ell_{\rm c}/v_{\rm A}^2 \sim 10^4 \,\omega_{\rm p}^{-1}$ indeed exceeds the time needed for the plasma to cross the domain. Note that the spectrum of S2b coincides with that of S2 at $t=5940\,\omega_{\rm p}^{-1}$, because the plasma has then just hit the reflective wall, and the shock has not formed yet.

\subsection{Particle acceleration} 
Standard theory depicts acceleration at a magnetized shock front as the result of 
diffusion back and forth across the shock, or of shock-drift motion along the mean advected electric field \citep{87Blandford}. In the relativistic limit, shock acceleration is ineffective~\cite{Begelman_1990,Lemoine_2006} unless intense turbulence can unlock particles off the field lines~\cite{Lemoine_2010}, and thereby trigger diffusive-type acceleration~\cite{Gallant+_99,Kirk+_00, Achterberg+_01}, a form of shock-drift process~\cite{Takamoto&Kirk_2015, Matsumoto17}, or a combination of both, meaning orbits in the regular field upstream of the shock, diffusive orbits downstream~\cite{2009MNRAS.393..587P}.
Let us stress here that what we mean by ``diffusive-type'' does not correspond to the standard ``spatial diffusion'' at play in subrelativistic shocks, but rather to ``diffusion in pitch-angle'', which ensures that particles can return to the shock. It is indeed known that in relativistic shocks, spatial diffusion does not have time to set in properly, because of the high advection velocity downstream of the shock and because the upstream particles can be caught back by the shock front just by barely changing their propagation direction~\cite{Gallant+_99, Achterberg+_01}. Similarly, the shock-drift type process that we will refer to in the following is sustained by pitch-angle scattering in the turbulence, which allows particles to remain close to the shock surface. It could be termed ``stochastic shock-drift'' in analogy with Ref.~\cite{Matsumoto17}, although those authors considered a subluminal configuration, not a superluminal one as in the present case.

To probe the acceleration process at work, we have tracked a large number $\sim \mathcal{O}( 10^5)$ of particles sampled in various (initial) energy intervals (see Appendix \ref{sec:particle_tracking}). Figure~\ref{fig: ptest} shows the trajectories and energy histories of four particles in S2, representative of the population able to circulate around the shock for an extended period of time. The Lorentz factor of some particles (e.g. orange and cyan in that figure) undergoes sizable oscillations before they encounter the shock; this results from their gyromotion along the fast-moving magnetic field lines~\cite{Wong+_20}, not from acceleration \emph{per se}. Notwithstanding this effect, the energization of the particles traveling in the vicinity of the shock is evident.  

We discriminate the acceleration processes in simulations S1 and S2 using the following argument. In a shock-drift process, the energy gained by a particle of velocity $\boldsymbol{v}$ equals the amount of work performed by the mean motional electric field $\boldsymbol{E_0} = v_\infty B_0\,\boldsymbol{\hat y}$, \emph{i.e.}, $W(E_0)=q\int{\rm d}t\, \boldsymbol{v}\cdot\boldsymbol{E_0}$, whereas for diffusive-type acceleration, the energy gain is rather related to the work $W(\delta E_z)= q\int {\rm d}t\, v_z \delta E_z$ performed by the motional turbulent electric field. The latter is mostly directed along $\boldsymbol{\hat z}$ because the plasma, which flows along $-\boldsymbol{\hat x}$, carries essentially $(\delta B_x,\,\delta B_y)$ magnetic fluctuations; hence $\delta E_z \simeq - v_\infty \delta B_y$. For each tracked particle with initial Lorentz factor at the onset of the powerlaw tail, {\it i.e.} $\gamma \geq 20$ (in S1) and $\gamma\geq200$ (S2), we have thus recorded $W(E_0)$ and $W(\delta E_z)$ during the time interval $\Delta t_{\rm sh}$ between the first and last encounters of the particle with the shock front. 

\begin{figure}
        \includegraphics[width=0.47\textwidth]{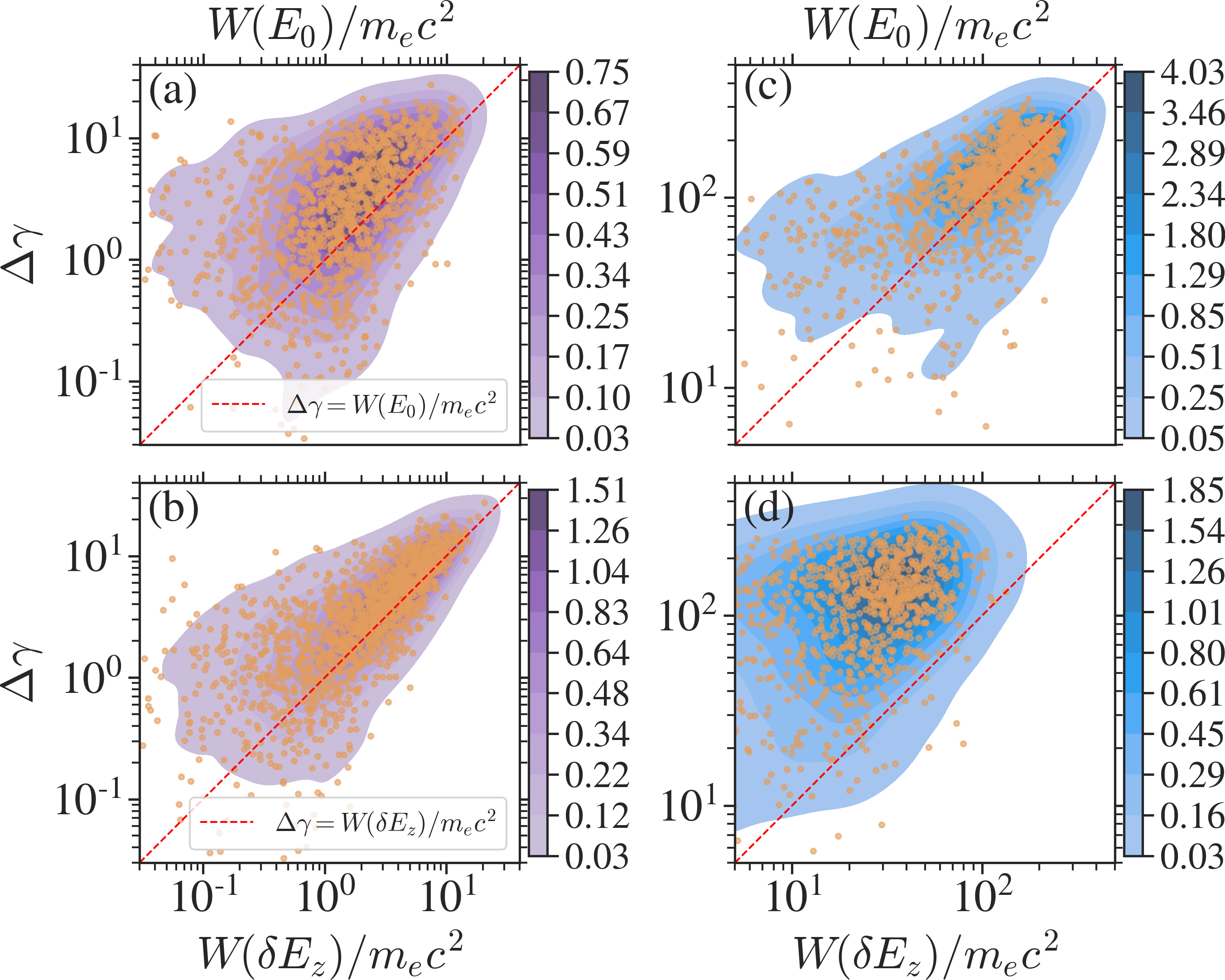}
    \caption{Correlation between the measured variation in Lorentz factor $\Delta\gamma$ and that predicted by shock-drift acceleration [$W(E_0)/m_{\rm e} c^2$, top row (a) and (c)] or diffusive-type acceleration [$W(\delta E_z)/m_{\rm e} c^2$, bottom row (b) and (d)] in simulations S1 [left column, (a) and (b)] and S2 [right column, (c) and (d)]. Orange dots represent individual measurements which are interpolated by the pseudocolor density plot.)
    }
    \label{fig: dgammavsdy}
\end{figure} 

Figure~\ref{fig: dgammavsdy} shows the correlations between the observed variation in Lorentz factor $\Delta \gamma$ during $\Delta t_{\rm sh}$ and $W(E_0)/m_{\rm e} c^2$ [top row, (a) and (c)], as well as $W(\delta E_z)/m_e c^2$ [bottom row, (b) and (d)], for simulations S1 [left column, (a) and (b)] and S2 [right column, (c) and (d)]. In these plots, the dashed red line indicates the expected level of contribution from either shock-drift or diffusive-type acceleration. Figures~\ref{fig: dgammavsdy}~(c) and (d) reveal that a shock-drift process sustained by particle scattering along the shock front nicely accounts for particle energization in S2, whereas in S1, acceleration appears dominated by diffusive-type acceleration. 


This general picture is further supported by the angular distribution of the suprathermal particle momenta (see Appendix \ref{sec:particle_tracking}). In S2, this angular map presents a clear asymmetry between positrons and electrons, roughly polarized along $\boldsymbol{E_0}$, such that positrons (resp. electrons) appear to drift with negative (resp. positive) $p_y$, as expected for $E_0 < 0$. By contrast, the angular map is significantly more isotropic in S1, as expected if diffusive-type acceleration dominates. Additionally, simulation S2 displays a net linear correlation between $\Delta \gamma$ and $\Delta t_{\rm sh}$, from which one can infer an acceleration timescale, $t_{\rm acc} \equiv \vert\langle\Delta\gamma/\gamma\rangle\vert ^{-1}\Delta t_{\rm sh} \sim 2-3\, p/(e E_0)$, again consistent with that expected for particles drifting along $\boldsymbol{E_0}$ at mildly relativistic speeds. 

The difference in spectral index observed between simulations S1 ($s\sim 2.5$) and S2 ($s\sim 3.5$) likely results from the distinct acceleration mechanisms at play. In particular, the spectral index for shock-drift acceleration -- at least, in subrelativistic shocks -- depends sensitively on the shock speed and on the ratio of the scattering frequency of particles to their gyrofrequency in the mean field~\citep{Takamoto&Kirk_2015}. At large scattering frequencies, the index approaches the canonical value of $2$, associated with diffusive shock acceleration, whereas at small scattering frequencies, the spectrum steepens significantly, encompassing the value measured in S2. 

Regarding the different acceleration processes in S1 and S2, we observe that the $r_{\rm g}$ vs $\ell_{\rm c}$ ordering, which controls the scattering rate of particles, varies between those two simulations because of different initial temperatures and magnetizations: $r_{\rm g}/ \ell_{\rm c} \sim 0.1$ at the onset of the powerlaw tail for S1, whereas $r_{\rm g} /\ell_{\rm c} \sim 1$ for S2. A detailed study of the ancillary simulation S4, which shares the same initial temperature as S1 and same magnetization as S2, reveals, however, that 
although a powerlaw index similar to that in S2 is obtained,
shock-drift and diffusive-type processes now contribute in about equal amounts to particle energization; accordingly, the angular map is less anisotropic than in S2, more than in S1. Overall, this suggests that both plasma temperature and magnetization may influence the prevalence of one mechanism over the other. While a comprehensive explanation for this change of regime certainly deserves further investigation, we emphasize that the main result of the present work, namely, the formation of a nonthermal spectrum at relativistic, magnetized turbulent shocks, is a robust feature.

Let us finally address simulation S3, characterized by a relativistically hot initial plasma and a substantial magnetization $\sigma_{\delta B}\sim 0.1$ (Appendix \ref{sec:sim_param_turb_scheme}). This simulation probes a new regime in which stochastic Fermi acceleration inside the pre-shock turbulence controls the acceleration process, because the (stochastic) acceleration timescale $t_{\rm acc} \sim \gamma_\infty \ell_{\rm c}/\sigma_{\delta B}c$ now becomes short enough ($\simeq 10^3\,\omega_{\rm p}^{-1}$) to energize the freshly injected particles before they attain the shock. Accordingly, the spectrum shown in Fig.~\ref{fig:spectrum_evolution} (lower panel) does not vary with time because the turbulence inside the simulation box is stationary, up to its fluctuations. This figure also reveals that the particle distribution has undergone significant heating beyond the simple shock-crossing conditions, as can be seen by direct comparison to simulation S2 (middle panel). Furthermore, we have verified that the same simulation, albeit with open boundary conditions to prevent shock formation, yields a similar spectrum. Finally, the spectral index $s \sim 3.5$ falls in line with that observed in PIC simulations of turbulence in the semi-relativistic regime~\cite{Zhdankin+_17, CS_18, Bresci+22}. 

\section{Discussion} \label{sec:conclusions}

Interestingly, the range of magnetizations that we consider here, $\sigma \sim 0.01\rightarrow 0.1$, and the range of spectral indices that we measure, $s\sim 2.5 \rightarrow 3.5$, appear rather typical of what is inferred from one-zone models of blazars~\citep{Celotti-Ghisellini_08} and gamma-ray bursts~\citep{Burgess+_2020}. This supports the idea that mildly relativistic shocks interacting with magnetized turbulence can play a leading role in dissipation and particle acceleration in a broad range of relativistic high-energy sources, up to moderate magnetizations. Our study thus significantly extends the realm where relativistic shock acceleration can operate without turbulence (\emph{i.e.} $\sigma\ll 10^{-4}$). As stochastic turbulent acceleration  is observed to take over shock acceleration at $\sigma\gtrsim 0.1$, one is tempted to sketch a picture in which, as the magnetization level rises, a shock, or a shock plus turbulence, then turbulence and eventually magnetic reconnection control dissipation and acceleration.

As noted, the present simulations are computationally expensive, which limits a broad parameter study. Future works should explore a larger parameter range, in particular larger dimensions (and dimensionalities) in order to examine how the particle spectrum changes with increasing $\ell_{\rm c}$, to make better contact with phenomenology.
\bigskip


\begin{acknowledgments}
We thank A. Bykov for insightful discussions. This work has been supported by the Sorbonne Universit\'e DIWINE Emergence-2019 program and by the ANR (UnRIP project, Grant No.~ANR-20-CE30-0030). V.~B. acknowledges support by the European Research Council under ERC-AdG Grant  No. PICOGAL-101019746. This work was granted access to the HPC resources of TGCC under the
allocations 2019-A0050407666, 2020-A0080411422, 2021-A0080411422 and 2022-A0130512993 made by GENCI. We thank X. Davoine for his assistance on particle-tracking diagnostics.
\end{acknowledgments}


\appendix


\section{Simulation parameters and turbulence generation scheme}
\label{sec:sim_param_turb_scheme}

Our PIC simulations are conducted with the fully electromagnetic and relativistic \textsc{calder} code \cite{Calder} which has been optimized to expunge beam-grid numerical instabilities, known to affect relativistic shock simulations \citep{Lemoine_2019, Vanthieghem_22}. We adopt a 2D3V geometry, with periodic boundary conditions in the transverse direction. In the longitudinal direction, particles are continually injected with mean velocity $\boldsymbol{v_\infty}=-0.87c\, \boldsymbol{\hat{x}}$ through the right-hand side boundary. At the left-hand boundary, conditions are either open or reflective for fields and particles, as discussed below. The mesh size is $\Delta x = \Delta y = 0.1 \,c/\omega_{\rm p}$, where $\omega_{\rm p} \equiv (4 \pi n_\infty e^2/m_{\rm e})^{1/2}$ represents the nonrelativistic plasma frequency of the far-upstream (injected) pair plasma, with $n_\infty= 2n_0/\gamma_\infty$ as the total proper density, $n_0$ the apparent density of one species and $\gamma_\infty = (1-v_\infty^2/c^2)^{-1/2}$. The simulation domain has dimensions of $48\,000\,\Delta x \times 6\,000\,\Delta y$. The time step is $\Delta t = 0.99\,\Delta x/c$. A uniform magnetic guide field $\boldsymbol{B_0}$ is applied along the (out-of-plane) $z$ direction. 
Each species (electrons or positrons) of the drifting plasma is initially represented by 10 particles per cell. 

Immediately after injection, the drifting plasma is subject to turbulence forcing in its proper frame via a Langevin antenna scheme~\cite{TenBarge_2014}. In detail, an external random current $j_{z,\rm ext} = (c/4\pi) \nabla^2 A_z$ with $A_z = \sum_{i=1}^{N_{\rm w}} a_i(t') e^{i \mathbf{k'_i \cdot r'}}$ excites external magnetic perturbations $\delta B_x$ and $\delta B_y$. The coefficients $a_i(t')$ obey the equation of a stochastically driven, damped harmonic oscillator. We use $N_{\rm w}=24$ plane waves, with mean wavenumber $\langle k' \rangle \simeq 2.9 \times 2 \pi /L_y$ (primed quantities are measured in the comoving plasma frame and $L_y = 600\,c/\omega_{\rm p}$ denotes the transverse box size). Numerically, the excitation scheme is implemented so as to have it evolved in the simulation grid, while the antenna external vector potential is evaluated on refined comoving grids 
\begin{align}
    t'&=\gamma_{\infty}\left[t-v_{\infty}(x-x_{\rm max})/c^2\right] \, , \label{eq: t turb proper frame} \\
    x'&=\gamma_{\infty}\left[-v_\infty  t+(x-x_{\rm max})\right],
    \label{eq: x turb proper frame}
\end{align}
with $x_{\rm max} = 4\,800\,c/\omega_{\rm p}$ the right-hand boundary of the domain, where particles are injected. While the choice of $\boldsymbol{\delta B}\cdot \boldsymbol{B_0}=0$ and $\boldsymbol{k} \cdot \boldsymbol{B_0}=0$ points to the excitation of Alfv\'en modes, we stress that no velocity perturbations are excited externally. Furthermore, the fluctuations are initialized with a large amplitude, which places them in the nonlinear regime. The role of this stirring is thus to initialize the system off-pressure balance so that it evolves rapidly towards a turbulent state. Consequently, we expect the resulting turbulence cascade to comprise a significant fraction of compressive modes, as discussed in Ref.~\cite{Zhdankin+_17}. 
By construction, our reduced dimensionality excites an anisotropic turbulence with $k_\parallel \ll k$, where $k_\parallel=k_z$ denotes the wavenumber component parallel to the mean field. Clearly, however, 3D simulations of the shock-turbulence interaction problem that we study remain prohibitive at the present time. Nevertheless, 3D and 2D simulations with an out-of-plane magnetic field have been shown to share the same characteristic in terms of turbulent cascade and particle acceleration, provided the turbulence level is large enough, as is the case here~\cite{CS_19, Bresci+22}.

\begin{table*}
\caption{\label{tab:param} Parameters defining the numerical simulations}
\begin{ruledtabular}
\begin{tabular}{lcccccc}
\textrm{Simulations}&
\textrm{S1}&
\textrm{S2}&
\textrm{S2a}&
\textrm{S2b}&
\textrm{S3}&
\textrm{S4}\\
\colrule
$\sigma_0$ & $0.2\times 10^{-3}$ & $0.6\times 10^{-3}$ & $0.6\times 10^{-3}$ & 
            $0.6\times 10^{-3}$ &  $0.6\times 10^{-3}$ &  $0.6\times 10^{-3}$ \\
$\sigma_{\delta B}$ & $2.\times 10^{-2}$ & $1.\times 10^{-2}$ & $0$ & 
            $1.\times 10^{-2}$ & $0.1$ &  $2.\times 10^{-2}$ \\
$k_B T/m_ec^2$ & $0.1$ & $4.$ & $4.$ & 
            $4.$ &  $4.$ &  $0.1$ \\
$eB_0/m_{\rm e} c \omega_{\rm p}$ & 0.03 & 0.2 & 0.2 & 0.2 & 0.38 & 0.05 \\
$e\delta B/m_{\rm e} c \omega_{\rm p}$ & 0.3 & 0.8 & $0.$ & 0.8 & 5. & 0.3 
\end{tabular}
\end{ruledtabular}
\end{table*}

Turbulence is excited over a few coherence lengths $\ell_{\rm c}$ in the vicinity of the right-hand boundary, $x_{\rm max}-600c/\omega_{\rm p} \leq x \leq x_{\rm max}$. Initially, the left-hand boundary ($x=0$) is left open to let the plasma exit freely. For all simulations but S2b, this boundary condition is turned to reflective once the turbulent part of the plasma has crossed the domain. This triggers a turbulent shock that mimics the interaction of two similar turbulent flows. As such, once the shock forms in the box, it propagates at a roughly constant speed.

The numerical parameters characterizing our reference simulations S1, S2 and S3 and the ancillary ones S2a (no turbulence), S2b (no shock) and S4 are compiled in Table~\ref{tab:param}. For S2b, the boundary conditions are left open on the left-hand side at all times, so that the turbulent plasma can exit freely the simulation domain, which prevents shock formation. 

We have also carried out additional runs in which the shock is triggered right at the onset of the simulation, in order to investigate the interaction of a shock with an ambient plasma that progressively becomes turbulent as time passes. While we do not observe a significant difference in terms of particle acceleration, the shock evolves in time, as it first interacts with a nonturbulent plasma, then with a turbulent plasma, which itself evolves downstream as it progressively fills the whole post-shock region.

\section{Particle-based diagnostics}
\label{sec:particle_tracking}

Regarding particle tracking, we follow a large number ($\sim 2 \times 10^5$) of particles, injected at different times and locations upstream of the shock, in order to study different histories of interaction with the upstream turbulence and the shock. We note that only a small fraction ($\lesssim 1\,$\%) of the tracked particles return to the shock after bouncing specularly on the reflective wall; such orbits do not therefore alter our results.

\begin{figure}
\begin{centering}
       \includegraphics[width=0.4\textwidth]{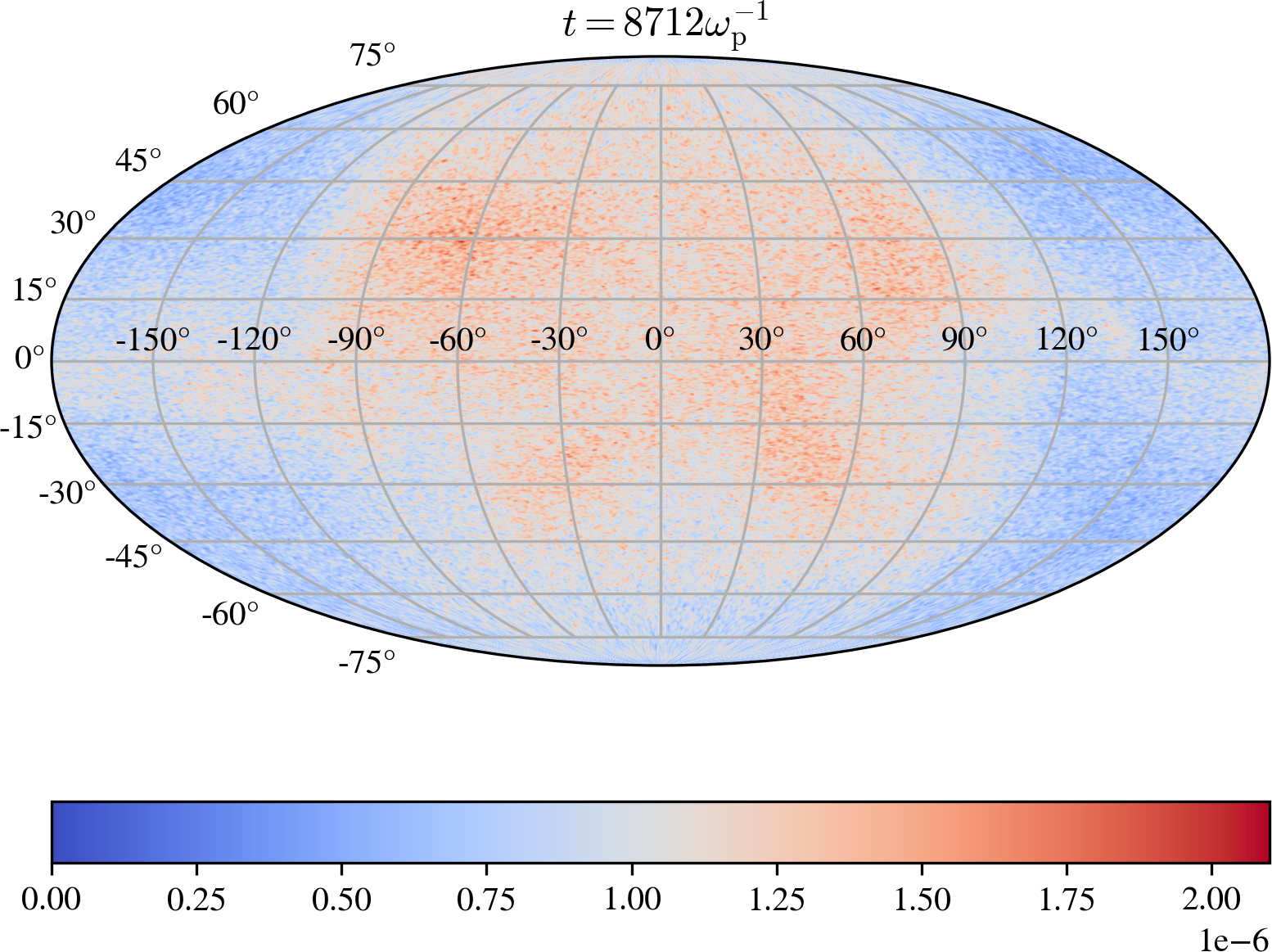}
       \includegraphics[width=0.4\textwidth]{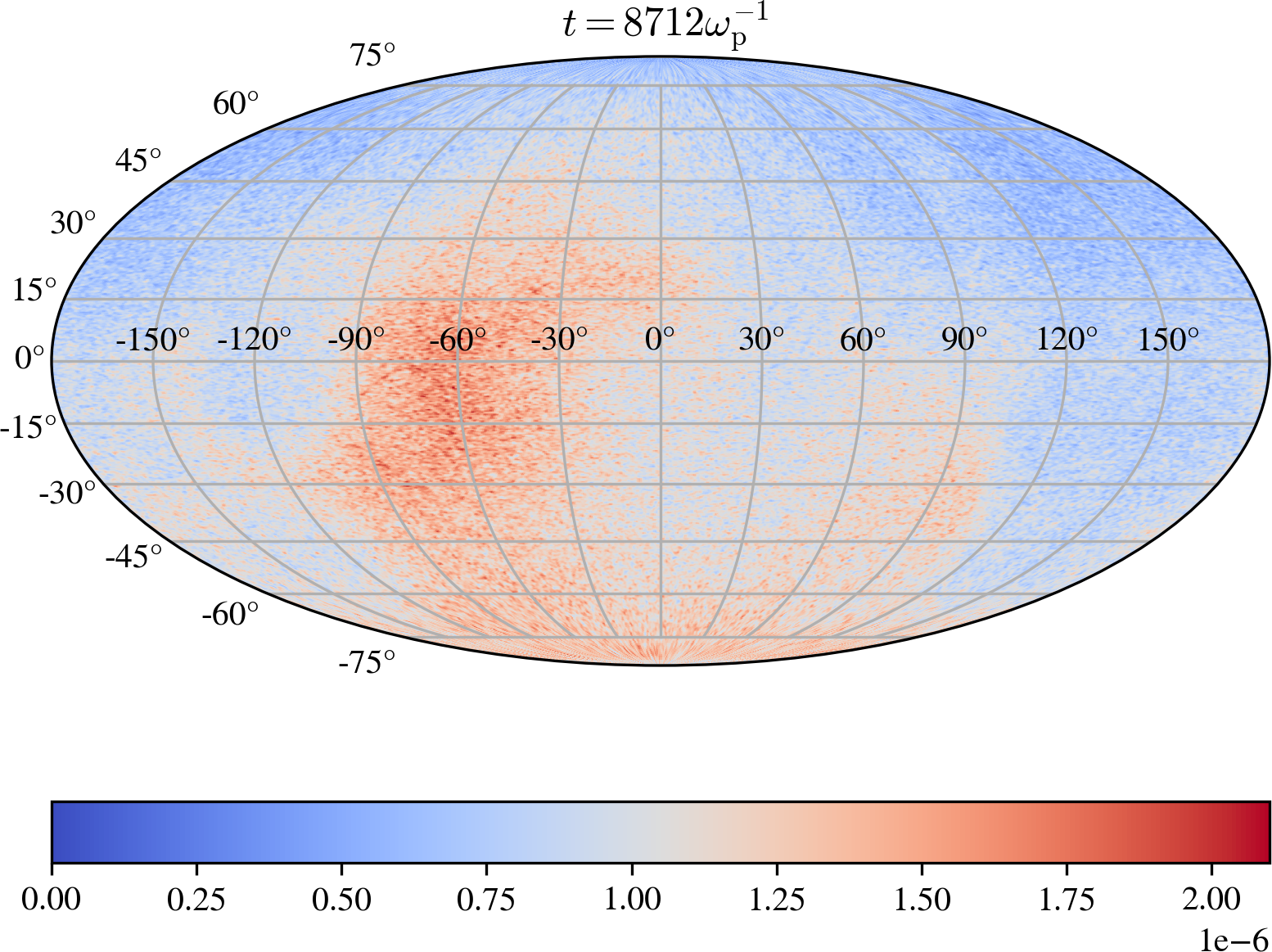}
      \caption{Normalized angular distributions of (left) positrons and (right) electrons, as obtained in simulation S1 at $t=8\,700\omega_{\rm p}^{-1}$, for particles with Lorentz factor $\gamma \geq 20$; see text for further details.
      }
    \label{fig: anisotropy_cold}
\end{centering}
\end{figure}

\begin{figure}
\begin{centering}
       \includegraphics[width=0.4\textwidth]{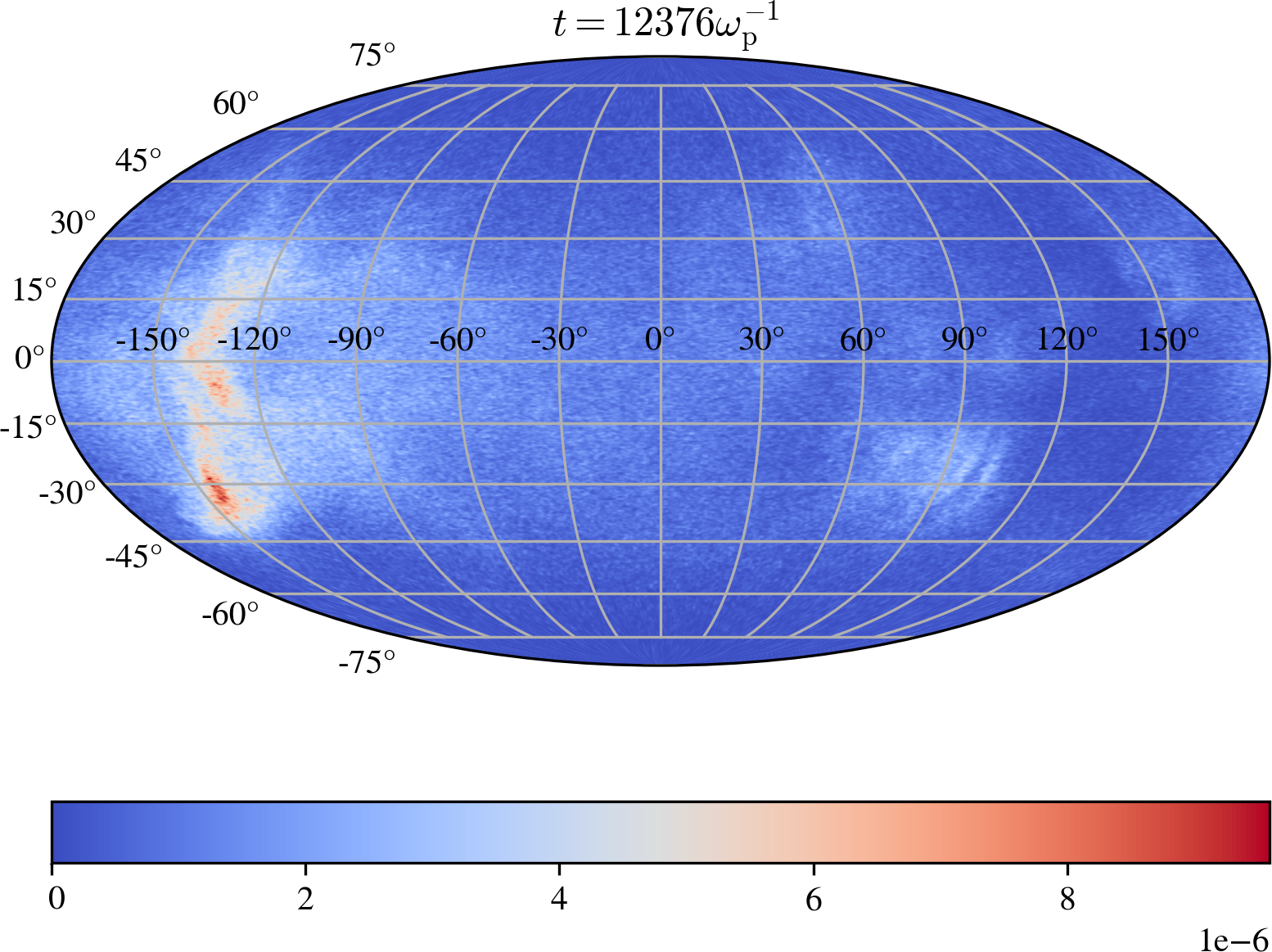}
       \includegraphics[width=0.4\textwidth]{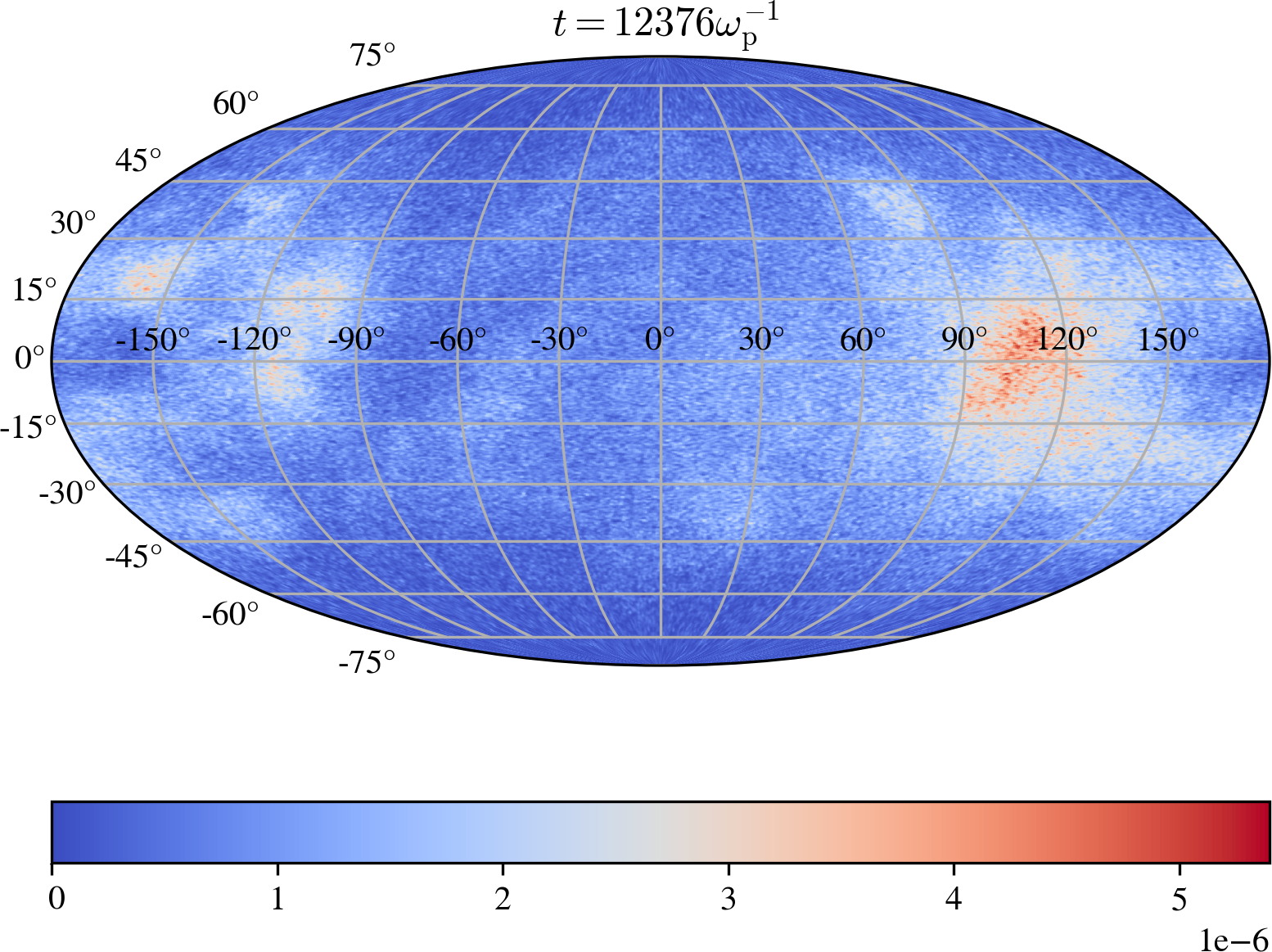}
      \caption{Same as Fig.~\ref{fig: anisotropy_cold}, for simulation S2 at $t=12\,400\, \omega_{\rm p}^{-1}$, for particles with Lorentz factors $\gamma \geq 200$; see text for further details.
      }
    \label{fig: anisotropy_hot}
\end{centering}
\end{figure}

We extract the mean shock velocity from a $x-t$ diagram of the plasma density in order to assign an analytical form to the shock front trajectory, which is thus averaged over the transverse dimension of the simulation box.
The amount of time ($\Delta t_{\rm sh}$) spent by each particle in the vicinity of the shock is obtained by comparing the trajectory of the particle with that of the shock front and locating the first and last shock-crossing times.
In Fig.~4 of the main text, we present the correlation obtained between the energy gained ($\Delta \gamma$) by the tracked particles during $\Delta t_{\rm sh}$ with the theoretical estimate for shock-drift acceleration, namely, $W(E_0) \equiv q E_0 \Delta y$, in terms of the mean electric field $\boldsymbol{E_0} = v_\infty B_0 \boldsymbol{\hat{y}}$ and $\Delta y$ the displacement along $y$. We also provide a correlation with the work performed by the motional electric fields carried by the incoming turbulent fluctuations drifting at the mean bulk velocity, $\boldsymbol{\delta E} \simeq - v_\infty \delta B_y \boldsymbol{\hat z}$. These electric fluctuations are mostly directed along $\boldsymbol{\hat z}$, because in our 2D3V simulations, the magnetic fluctuations mostly lie in the simulation plane while the plasma drifts along $\boldsymbol{\hat x}$. 
$W(E_0)$ accounts for shock-drift acceleration, while $W(\delta E_z)$ characterizes diffusive-type Fermi acceleration. The latter mechanism must be distinguished, of course, from purely stochastic acceleration, for which the contributing electric field scales as $-\boldsymbol{\delta v}\times\boldsymbol{\delta B}$, with $\boldsymbol{\delta v}$ the turbulent velocity component as measured in the upstream rest frame; those provide smaller contributions, as we have checked. The amount of work performed by the various electric fields is measured over the time interval $\Delta t_{\rm sh}$ between the first and last shock crossings for each particle. The correlation is then built from the sample obtained. In simulation S1, the particles that we track have Lorentz factors $\gamma \geq 20$, while in S2, $\gamma\geq 200$.

Finally, as discussed in the main text, we have also recorded angular distributions of the accelerated particles to probe a possible anisotropy associated with the drift along the mean motional electric field $\boldsymbol{E_0} \parallel \boldsymbol{\hat y}$. The angular distribution has been extracted  at the final time of the simulation from the whole sample of particles (not only the tracked ones) that lie within $\pm250\,c/\omega_{\rm p}$ of the shock front, and whose Lorentz factor $\gamma > 20$ for simulation S1, and $\gamma>200$ for simulation S2, {\it i.e.} in the powerlaw tail. Those angular maps are displayed in Fig.~\ref{fig: anisotropy_cold} for simulation S1 and Fig.~\ref{fig: anisotropy_hot} for S2, with for each a map for positrons (left) and one for electrons (right). The longitude is here defined as $\phi \equiv \mathrm{arctan2} \left(p_y, p_x\right) \in [-\pi, \pi]$ and the latitude as $\theta \equiv \pi/2 - \arccos\left(p_z/p\right) \in[-\pi/2, \pi/2]$, in terms of the momentum components $(p_x,\,p_y,\,p_z)$ and norm $p$. 
As discussed in the main text, the positron and electron angular distributions are strongly anisotropic in S2, with bright peaks at approximately opposite angles relative to the shock normal. In detail, positrons (electrons) are preferentially drifting with $p_y < 0$ (resp. $p_y >0$), consistent with shock-drift acceleration (noting that $E_0 < 0$ here, since $B_0 >0$ and $v_\infty <0$). In S1, the suprathermal particles show a much broader angular distribution with weaker differences between electrons and positrons, consistent with the dominant mechanism of diffusive-type acceleration.

\end{document}